\documentclass[reqno,12pt]{amsart}

\newtheorem{prop}{Proposition}

\newcommand\eps\varepsilon
\newcommand\ph\varphi
\newcommand\kap\Lambda

\newcommand \bs\boldsymbol

\usepackage{enumerate}
\usepackage{graphicx}
\usepackage{float}

\begin{document}

\title[System with Anisotropic Dry Friction]
{On a Strange Motion in the  System with Anisotropic Dry Friction   }

\author[Oleg Zubelevich]{Oleg Zubelevich\\ \\\tt
 Steklov Mathematical Institute of Russian Academy of Sciences\\
 \\Dept. of Theoretical mechanics,  \\
Mechanics and Mathematics Faculty,\\
M. V. Lomonosov Moscow State University\\
Russia, 119899, Moscow,  MGU \\ozubel@yandex.ru
 }
\email{ozubel@yandex.ru}
\date{}
\thanks{The research was funded by a grant from the Russian Science
Foundation (Project No. 19-71-30012)}

\begin{abstract}We consider a particle on the horizontal plane with a dry friction. The friction is anisotropic but symmetric under the group of rotations of the plane. 

It turns out that the particle can move such that in the finite time the trajectory turns about the center of symmetry  infinitely many times.  
\end{abstract}

\maketitle
\numberwithin{equation}{section}
\newtheorem{theorem}{Theorem}[section]
\newtheorem{lemma}[theorem]{Lemma}
\newtheorem{definition}{Definition}[section]

\section{Introduction and the Statement of The Problem}
Consider a fixed horizontal table. There is a particle of mass $m$ on the table. There is an anisotropic  dry friction between the particle and the table. 

To explain how this friction acts on the particle let us introduce the standard polar coordinates $(r,\ph)$   in the plane.
And let $O$ be the origin.

Define the friction force  by the formula
\begin{equation}\label{sg678}
\bs F=-kmg\frac{(\bs v,\bs e_\ph)}{|\bs v|}\bs e_\ph,\quad |\bs v|\ne 0.\end{equation}
Here $(\cdot,\cdot),\quad |\cdot|$ stand for the inner product and the  length of a vector respectively; $\bs e_\ph$ stands for the coordinate unit vector which is directed along the coordinate circle; $\bs v$ stands for a velocity of the particle and $k>0$ is a constant friction coefficient. 

Thus formula (\ref{sg678}) means that the friction is directed along  the concentric circles with the common center $O$. 

Such a hypothesis is consistent with the modern notion on anisotropic dry friction \cite{z1,z2,d,k}.  

From the results of these articles we know that a good model for the force of  anisotropic dry friction is as follows:
$$\bs F_{\mathrm{friction}}=-NA(\bs r)\frac{\bs v}{|\bs v|}.$$ Here $A$ is a linear operator (matrix) which depends on the position vector
$\bs r$ of the particle; $N$ -- normal pressure. 

The operator $A$ is non negatively defined: $(A(\bs r)\bs v,\bs v)\ge 0,\quad \forall \bs r,\bs v.$

\section{The Main Observation}
In the polar coordinates the Second Newton Law $m\bs{\dot v}=\bs F$ with (\ref{sg678}) takes the form
\begin{equation}\label{sfvvv}
\ddot r=r\dot\ph^2,\quad r\ddot\ph+2\dot r\dot\ph=-\sigma\frac{r\dot\ph}{\sqrt{\dot r^2+(r\dot\ph)^2}},\quad \sigma=kg.\end{equation}
This system is considered under the assumption $\dot r^2+(r\dot\ph)^2\ne 0$.

The following proposition is obtained by direct calculation.
\begin{prop}\label{dfgtyu}
System (\ref{sfvvv}) has a solution
\begin{equation}\label{dgh67}r(t)=\frac{\sigma}{3\sqrt 6}(T-t)^2,\quad \ph(t)=\pm\sqrt 2\ln(T-t)+K,\quad t\in(-\infty, T).
\end{equation}
Here  $K,T\in\mathbb{R}$ are arbitrary constants.\end{prop}
This solution starts from a moment $t_0<T$ and comes to the origin $O$ in time $T-t_0$. During this time the particle turns around the origin infinitely many times.   

Note also that  solution (\ref{dgh67}) describes the Logarithmic spiral in the plane.

\end{document}